\documentclass[english,twocolumn,10pt,Engish,superscriptaddress,nofootbible]{revtex4-1}
\usepackage[T1]{fontenc}
\usepackage[latin9]{inputenc}
\usepackage{color}
\usepackage{babel}
\usepackage{array}
\usepackage{verbatim}
\usepackage{multirow}
\usepackage{amsmath}
\usepackage{graphicx}
\usepackage{esint}
\usepackage[unicode=true,pdfusetitle,
 bookmarks=true,bookmarksnumbered=false,bookmarksopen=false,
 breaklinks=false,pdfborder={0 0 1},backref=false,colorlinks=true]
 {hyperref}
\hypersetup{
 pdfborderstyle=,linkcolor=blue,citecolor=cyan}

\makeatletter

\providecommand{\tabularnewline}{\\}

\usepackage{babel}
\usepackage{xcolor}
\usepackage{amsmath}

\makeatother

\begin{document}
\title{Nuclear deformation effects in photoproduction of $\rho$ mesons in
ultraperipheral isobaric collisions }
\author{Shuo Lin}
\email{linshuo@mail.ustc.edu.cn}

\affiliation{Department of Modern Physics and Anhui Center for Fundamental Sciences in Theoretical Physics, University of Science and Technology
of China, Anhui 230026, China}
\author{Jin-Yu Hu}
\email{jinyuhu2000@mail.ustc.edu.cn}

\affiliation{Department of Modern Physics and Anhui Center for Fundamental Sciences in Theoretical Physics, University of Science and Technology
of China, Anhui 230026, China}
\author{Hao-Jie Xu}
\email{haojiexu@zjhu.edu.cn}

\affiliation{School of Science,Huzhou University,Huzhou, Zhejiang, 313000, China}
\affiliation{Strong-Coupling Physics International Research Laboratory (SPiRL),
Huzhou University, Huzhou, Zhejiang 313000, China}

\author{Shi Pu}
\email{shipu@ustc.edu.cn}
\affiliation{Department of Modern Physics and Anhui Center for Fundamental Sciences in Theoretical Physics, University of Science and Technology
of China, Anhui 230026, China}
\affiliation{Southern Center for Nuclear-Science Theory (SCNT), Institute of Modern
Physics, Chinese Academy of Sciences, Huizhou 516000, Guangdong Province,
China}
\affiliation{Shanghai Research Center for Theoretical Nuclear Physics, NSFC and Fudan University, Shanghai 200438, China}

\author{Qun Wang}
\email{qunwang@ustc.edu.cn}
\affiliation{Department of Modern Physics and Anhui Center for Fundamental Sciences in Theoretical Physics, University of Science and Technology
of China, Anhui 230026, China}
\affiliation{School of Mechanics and Physics, Anhui University of Science and Technology,
Huainan,Anhui 232001, China}
\begin{abstract}
We investigate the photoproduction of neutral $\rho$ mesons in ultraperipheral
isobaric collisions of $\ensuremath{_{44}^{96}}\textrm{Ru}+\ensuremath{_{44}^{96}}\textrm{Ru}$
and $\ensuremath{_{40}^{96}}\textrm{Zr}+\ensuremath{_{40}^{96}}\textrm{Zr}$
at $s_{NN}^{1/2}=200$ GeV, employing the dipole model within the
equivalent photon approximation. By implementing the Woods-Saxon distribution
for the nuclear mass density, which is derived from the density functional
theory incorporting nuclear deformation effects, we calculate differential
cross sections in Ru$+$Ru and Zr$+$Zr collisions as well as their
ratios in the transverse momentum spectrum squared $q_{T}^{2}$. We
observe the characteristic dip behavior in differential cross sections,
an indicator of diffraction in high-energy collisions. When nuclear
deformation is considered, the ratio of differential cross sections
shows a nearly linear increase behavior with increasing $q_{T}^{2}$
for $q_{T}^{2}\lesssim0.015$ $\textrm{GeV}^{2}$, in agreement with
experimental observation. We offer a simple explanation for such a
behavior by introducing an effective description of the thickness
function. We also extend our discussion on the photoproduction of
neutral $\rho$ mesons in electron-ion collisions with nuclear targets
Cu-63, Au-197 and U-238.
\end{abstract}
\maketitle

\section{Introduction}

Relativistic heavy-ion collisions (HIC) provide a platform for exploring
novel phenomena under extreme conditions, such as quantum transport
phenomena associated with chiral anomaly \citep{Kharzeev:2007jp,Fukushima:2008xe,Huang:2013iia,Pu:2014cwa,Chen:2016xtg,Hidaka:2017auj}
(also see recent reviews \citep{Gao:2020vbh,Gao:2020pfu,Kharzeev:2015znc,Hidaka:2022dmn})
in ultra-strong electromagnetic fields \citep{Skokov:2009qp,Bzdak:2011yy,Deng:2012pc,Roy:2015coa},
and nonlinear effects of quantum electrodynamics (QED) \citep{Hattori:2012je,Hattori:2012ny,Hattori:2020htm,STAR:2019wlg,Hattori:2022uzp,Adler:1971wn,Schwinger:1951nm,Copinger:2018ftr,Copinger:2020nyx,Copinger:2022jgg}.
Furthermore, such strong electromagnetic fields in HIC, regarded as
a huge flux of quasi-real photons, offer a new opportunity to investigate
photon-photon and photon-nuclear interactions, which are significantly
enhanced by proton numbers of colliding nuclei. 

Dilepton photoproduction, one of the key processes in photon-photon
interactions, has been extensively investigated in ultraperipheral
collisions (UPC) \citep{ATLAS:2018pfw,STAR:2018ldd,STAR:2019wlg,ALICE:2022hvk}
using the equivalent photon approximation (EPA) \citep{Klein:2016yzr,Bertulani:1987tz,Bertulani:2005ru,Baltz:2007kq}.
However it has been found that EPA cannot describe the data of dilepton
photoproduction in UPC accurately, primarily due to its omission of
the impact parameter and the transverse momentum dependence of initial
photons. Then the EPA has been extended to generalized EPA or QED
in the background field approach \citep{Vidovic:1992ik,Hencken:1994my,Hencken:2004td,Zha:2018tlq,Zha:2018ywo,Brandenburg:2020ozx,Brandenburg:2021lnj,Li:2019sin,Wang:2022ihj}
based on the factorization theorem \citep{Klein:2018fmp,Klein:2020jom,Li:2019yzy,Xiao:2020ddm}
and the QED model incorporating a wave-packet description of nuclei
\citep{Wang:2021kxm,Wang:2022gkd,Lin:2022flv}. To account for transverse
momentum broadening effects, higher-order contributions have been
considered, such as Coulomb corrections \citep{Ivanov:1998ka,Eichmann:1998eh,Segev:1997yz,Baltz:1998zb,Baltz:2001dp,Baltz:2003dy,Baltz:2009jk,Zha:2021jhf,Klein:2020jom,Sun:2020ygb}
arising from multiple re-scattering within intense electromagnetic
fields and the Sudakov effect pertaining to the soft photon radiation
in the final state \citep{Li:2019sin,Li:2019yzy,Klein:2018fmp,Klein:2020jom,Shao:2023zge,Shao:2022stc}. 

Vector meson photoproduction, a form of photon-nucleus interaction
in high-energy collisions, provide rich information for parton structures
inside nuclei and nucleons \citep{Guo:2021ibg,Guo:2023qgu,Sun:2021pyw,Koempel:2011rc,Brodsky:1994kf,Hatta:2019lxo,Mantysaari:2022ffw,Mantysaari:2023qsq,Arslandok:2023utm,Achenbach:2023pba}.
Compared to $ep$ or $eA$ collisions, UPC provide a strong photon
flux, thereby offer an exceptional opportunity to investigate photon-nucleus
processes. New data have been obtained in recent experiments \citep{ALICE:2019tqa,ALICE:2020ugp,ALICE:2021gpt,STAR:2019wlg,STAR:2021wwq}
for the study of gluon distributions \citep{Zhou:2022twz,Brandenburg:2022jgr,Hagiwara:2021xkf,Zha:2020cst}.
Recently, the high-energy version of the double-slit interference
phenomenon, a milestone in UPC physics, was observed in the process
$\gamma A\rightarrow\rho^{0}A$ and $\rho^{0}\rightarrow\pi^{+}\pi^{-}$
\citep{STAR:2022wfe}. Such an interference phenomenon was studied
using the color glass condensate effective theory \citep{Xing:2020hwh}
and the vector meson dominance model \citep{Zha:2020cst}. Meanwhile,
the asymmetry of $\pi^{+}\pi^{-}$ from $\rho^{0}$ decay was discussed
in Refs. \citep{Hagiwara:2020juc,Hagiwara:2021xkf,Xing:2020hwh},
and the $\cos(4\phi)$ azimuthal asymmetry was highlighted as a possible
signal for the elliptic gluon distribution \citep{Hagiwara:2021xkf}. 

While many studies have been conducted on ultraperipheral Au+Au or
Pb+Pb collisions, photon-photon and photon-nuclear interactions in
ultraperipheral isobaric collisions have not yet been systematically
discussed. Unlike spherical nuclei, Ru-96 and Zr-96 nuclei have deformation
in charge and density distributions. In a recent work by some of us
\citep{Lin:2022flv}, the distributions of the transverse momentum,
invariant mass, and azimuthal angle for dileptons in photoproduction
processes of isobaric collisions have been calculated. The main result
of the study suggests that nuclear structure can significantly influence
these spectra. A subsequent study about the nuclear deformation effect
on dilepton photoproduction is presented in Ref. \citep{Luo:2023syp}.
The excitation of Ru and Zr nuclei through photon-nuclear interaction
and neutron emission is discussed in Ref. \citep{Zhao:2022dac}. 

\begin{table*}
\caption{Parameters in WS distributions for Ru and Zr in (a) deformed case
and (b) spherical case, followed by the procedure in Refs. \citep{Xu:2021qjw,Xu:2021vpn}.
\protect\label{tab:Parameters}}

\centering%
\begin{tabular}{c|c|c|c|c}
\hline 
(a) with deformation & $R$ (fm) & $a$ (fm) & $\beta_{2}$ & $\beta_{3}$\tabularnewline
\hline 
Ru-96 & 5.093 & 0.471 & 0.16 & 0\tabularnewline
\hline 
Zr-96 & 5.021 & 0.517 & 0 & 0.20\tabularnewline
\hline 
\end{tabular}$\;\;\;\;$%
\begin{tabular}{c|c|c|c|c}
\hline 
(b) w/o deformation & $R$ (fm) & $a$ (fm) & $\beta_{2}$ & $\beta_{3}$\tabularnewline
\hline 
Ru-96 & 5.093 & 0.487 & 0 & 0\tabularnewline
\hline 
Zr-96 & 5.022 & 0.538 & 0 & 0\tabularnewline
\hline 
\end{tabular}
\end{table*}

In this work, we investigate the photoproduction of $\rho^{0}$ mesons
in ultraperipheral isobaric collisions using the dipole model with
EPA. We introduce deformation parameters into the mass density distribution
inspired by state-of-the-art calculations in density functional theory.
We will calculate transverse momentum spectra of $\rho^{0}$ mesons
by photoproduction in ultraperipheral isobaric $_{44}^{96}$Ru$+_{44}^{96}$Ru
and $_{40}^{96}$Zr$+_{40}^{96}$Zr collisions at $s_{NN}^{1/2}=$200
GeV. 
We will show when nuclear deformation is accounted for, the ratio of differential cross sections increase with increasing transverse momentum spectrum squared $q_{T}^{2}$, aligning with experimental observations. We will offer a simple explanation for this behavior through the introduction of an effective description of the thickness function.

The paper is organized as follows. In Sec. \ref{sec:Theoretical}
and \ref{sec:Phenomenology-setup}, we briefly introduce the dipole
model and phenomenological models and parameters for numerical calculations,
respectively. The numerical results on transverse momentum spectra
in isobaric collisions are presented in Sec. \ref{sec:distribution}.
We will discuss possible application to photon-nuclear interaction
at the Electron Ion Collider (EIC) in Sec. \ref{sec:EIC}. A summary
of the results is given in Sec. \ref{sec:Conclusion}. Throughout
this work, we adopt the convention for the metric tensor $g_{\mu\nu}=\textrm{diag}\{+,-,-,-\}$.

\section{Theoretical method }

\label{sec:Theoretical} We first outline the strategy of this work.
For a comprehensive analysis, one might follow the theoretical method
in Refs. \citep{Xing:2020hwh,Zha:2020cst}, which include the impact
parameter and the transverse momentum dependence of initial photons.
This approach aligns with recent studies on vector meson photoproduction
in heavy-ion collisions \citep{Mantysaari:2023prg} and fluctuating
nucleon substructure in $J/\psi$ photoproduction \citep{Mantysaari:2022sux}.
However, differential cross sections for these processes are rather
complicated, which presents a challenge to extract nuclear deformation
effects. To streamline our discussion, we implement the EPA, thereby
neglecting the impact parameter and transverse momentum dependence
of initial photons in our current study. We will demonstrate that
even with the EPA, it is possible to capture the essential physics
of nuclear deformation effects in vector meson photoproduction in
isobaric collisions.

The cross section for vector meson production in ultraperipheral heavy-ion
collisions can be written in the EPA (see, e.g., Ref. \citep{Bertulani:2005ru}
for a review of the topic). In the EPA, the vector meson production
can be understood 
as the result of a photon flux generated by one nucleus colliding with another nucleus, thereby creating vector mesons.
If we focus on the case
in which the photon flux is generated by the nucleus $A_{1}$, then
the cross section can be simply written as, 
\begin{eqnarray}
\sigma & = & \int d\omega\frac{n_{1}(\omega)}{\omega}\sigma_{\gamma A_{2}\rightarrow\rho A_{2}}(\omega),\label{eq:EPA_01}
\end{eqnarray}
where $\sigma_{\gamma A\rightarrow\rho A}$ is the cross section for
the photoproduction of the vector meson in photon-nucleus collisions
and $n_{1}(\omega)$ is the photon flux with energy $\omega$ generated
by $A_{1}$. By integration of the equivalent photon flux per unit
area over the impact parameter we have 
\begin{align}
n(\omega)= & \frac{2Z^{2}\alpha}{\pi\beta^{2}}\left\{ \xi K_{0}(\xi)K_{1}(\xi)\right.\nonumber \\
 & \left.-(\xi^{2}/2)\left[K_{0}^{2}(\xi)-K_{1}^{2}(\xi)\right]\right\} ,
\end{align}
where $K_{i}(\xi)$ are Bessel function of order $i$, and $\xi=2\omega R_{A}/(\gamma\beta)$
with $\alpha$, $\beta$, $\gamma$ and $Z$ being the fine-structure
constant, velocity and Lorentz contraction factor of the nucleus,
and nuclear charge number, respectively.

In realistic nucleus-nucleus collisions, both $A_{1}$ and $A_{2}$
can be sources of photon flux and then scatter with the alternative
nucleus. Therefore, the cross section in Eq. (\ref{eq:EPA_01}) should
be extended as, 
\begin{align}
\sigma= & \int dy[n_{1}(y)\sigma_{\gamma A_{2}\rightarrow\rho A_{2}}(y)\nonumber \\
 & +n_{2}(-y)\sigma_{\gamma A_{1}\rightarrow\rho A_{1}}(-y)],\label{eq:cross_section_02}
\end{align}
where $y=\ln(2\omega/m_{V})$ is the rapidity of the vector meson
with $m_{V}$ being the vector meson's mass. We emphasize that in
principle there should be interference terms in the amplitude squared,
see, e.g., Refs. \citep{Xing:2020hwh,Zha:2020cst,Mantysaari:2023prg,STAR:2022wfe}
for discussions on the\textcolor{red}{{} }quantum interference in relativistic
heavy-ion collisions. We will concentrate on the effects caused by
the nuclear structure in the current work, so we will neglect these
interference terms in the EPA since they are not essential to the
effects we are looking at. We will comment on possible contributions
from these interference terms at the end of the next section.

The dominant process for vector meson production in UPC is the coherent
one, and its differential cross section can be described by the dipole
model \citep{Brodsky:1994kf,Kowalski:2003hm,Kowalski:2006hc,Ryskin:1992ui},
\begin{eqnarray}
\frac{d\sigma^{\gamma A\rightarrow\rho A}}{dq_{T}^{2}} & = & \frac{1}{16\pi}\left|\mathcal{A}\right|^{2}.\label{eq:sub_cross_section}
\end{eqnarray}
The scattering amplitude $\mathcal{A}$ can be conventionally expressed
as the convolution of the dipole scattering amplitude $N(\mathbf{r}_{T},\mathbf{b}_{T})$,
the vector meson's wave function $\Psi^{V\rightarrow q\bar{q}}(r_{\bot},z)$,
and photon splitting functions $\Psi^{\gamma\rightarrow q\bar{q}}(r_{\bot},z)$
in position space,
\begin{align}
\mathcal{A}= & 2i\int d^{2}\mathbf{b}_{T}e^{i\mathbf{\Delta}_{T}\cdot\mathbf{b}_{T}}\int\frac{d^{2}\mathbf{r}_{T}}{4\pi}\int_{0}^{1}dz\nonumber \\
 & \times\Psi^{\gamma\rightarrow q\bar{q}}(\mathbf{r}_{T},z)\Psi^{V\rightarrow q\bar{q}*}(\mathbf{r}_{T},z)\nonumber \\
 & \times N(\mathbf{r}_{T},\mathbf{b}_{T}),
\end{align}
where '$*$' means complex conjugate, two dimensional $\mathbf{r}_{T}$
is the transverse size of the $q\overline{q}$ dipole, two dimensional
$\mathbf{b}_{T}$ is the impact parameter of the dipole relative to
the target center, $-\mathbf{\Delta}_{T}$ is the recoil transverse
momentum of the nucleus, and $z=k^{+}/p^{+}$, with $k^{+}$ being
the light cone momentum of the quark in the $q\overline{q}$ pair
and $p^{+}$ the light cone momentum of the photon. In the EPA, the
transverse momentum of the photon is negligible, thus the transverse
momentum of the vector meson $\mathbf{q}_{T}$ should be equal to
$\Delta_{T}$. Here, the dipole and nucleus scattering amplitude 
\begin{align}
N(\mathbf{r}_{T},\mathbf{b}_{T})= & 1-\frac{1}{N_{c}}\textrm{Tr}\left[U(\mathbf{b}_{T}+\mathbf{r}_{T}/2)\right.\nonumber \\
 & \left.U^{\dagger}(\mathbf{b}_{T}-\mathbf{r}_{T}/2)\right]
\end{align}
where $N_{c}$ is the number of colors and 
\begin{equation}
U(\mathbf{x}_{T})=\mathcal{P}\exp\left[ig\int_{-\infty}^{+\infty}dx^{+}A^{-}(x^{+},\mathbf{x}_{T})\right],
\end{equation}
is the Wilson line \citep{Gelis:2010nm,Berges:2020fwq} with $\mathcal{P}$
standing for the path ordering.

The photon splitting function $\Psi^{\gamma\rightarrow q\bar{q}}$ describes the  process by which a photon splits into quark-antiquark pairs. This function can be calculated using perturbative theories, e.g. see the textbook \citep{Kovchegov:2012mbw}.
In contrast, the vector meson wave function $\Psi^{V\rightarrow q\bar{q}*}$ is
nonperturbative and cannot be derived from first principle.  Consequently, we parameterize this function following the approach outlined in Ref. \citep{Kowalski:2006hc}. The product of the vector meson wave function and photon splitting functions is presented as follows,
\begin{align}
 & \Psi^{\gamma\rightarrow q\bar{q}}\Psi^{V\rightarrow q\bar{q}*}\nonumber \\
= & \frac{N_{c}ee_{q}}{\pi}\bigg\{ m_{q}^{2}K_{0}(|\mathbf{r}_{T}|\varepsilon_{f})\Phi^{*}(|\mathbf{r}_{T}|,z)\nonumber \\
 & +\frac{\partial K_{0}(|\mathbf{r}_{T}|\varepsilon_{f})}{|\partial\mathbf{r}_{T}|}\frac{\partial\Phi^{*}(|\mathbf{r}_{T}|,z)}{\partial|\mathbf{r}_{T}|}\left[z^{2}+(1-z)^{2}\right]\bigg\},
\end{align}
where $e_{q}$ is the quark's electric charge, the parameter $\varepsilon_{f}$
in our case is reduced to $\varepsilon_{f}\approx m_{q}$ and $\Phi^{*}(|\mathbf{r}_{T}|,z)$
is the scalar part of the vector meson's wave function,
\begin{eqnarray}
\Phi^{*}(|\mathbf{r}_{T}|,z) & = & C_{V}z(1-z)\exp\left(-\frac{|\mathbf{r}_{T}|^{2}}{R_{T}^{2}}\right),
\end{eqnarray}
with $C_{V}=4.47$ and $R_{T}^{2}=21.9\;\textrm{GeV}{}^{-2}$ for
$\rho$ meson. 
 In this work, the meson wave function employed is referred to as Light Cone Gaussian wave functions. It is noteworthy that alternative wave function, named the boosted Gaussian wave function, could also be utilized. This approach is extensively discussed in literature, see e.g. Refs.~\citep{SampaiodosSantos:2014qtt, Forshaw:2003ki, Goncalves:2014swa, Goncalves:2017wgg, Goncalves:2021ziy, Armesto:2014sma} and reference therein for details.
Additionally, Ref.~\cite{Goncalves:2020cir} suggests the introduction of a new factor for $\Psi^{\gamma\to q\bar{q}}(\mathbf{r}_{T}, z)$ due to some non-perturbative QCD corrections, by modifying it as follows:
$
\Psi^{\gamma\to q\bar{q}}(\mathbf{r}_{T}, z) \rightarrow f_{c}(\mathbf{r}_{T}) \Psi^{\gamma\to q\bar{q}}(\mathbf{r}_{T}, z),
$
which leads to a reduction in the cross section.

\section{Models and parameters }

\label{sec:Phenomenology-setup} In the previous section, we briefly
introduced the cross section for the vector meson production in UPCs
using the EPA. We note that the dipole scattering amplitude $N(\mathbf{r}_{T},\mathbf{b}_{T})$
is non-perturbtaive. In this work, we adopt the conventional parameterization
for $N(\mathbf{r}_{T},\mathbf{b}_{T})$ which is widely used in Refs.
\citep{Brodsky:1994kf,Kowalski:2003hm,Kowalski:2006hc,Ryskin:1992ui},
which is connected to the thickness function $T_{A}(\mathbf{b}_{T})$
by 
\begin{eqnarray}
N(\mathbf{r}_{T},\mathbf{b}_{T}) & = & 1-\exp\left[-2\pi B_{p}AT_{A}(\mathbf{b}_{T})\mathcal{N}(\mathbf{r}_{T})\right],\label{eq:dipole_amplitude}
\end{eqnarray}
where $A$ is the number of nucleons in colliding nuclei ($A_{1}$
or $A_{2}$), $B_{p}$ is a constant and $\mathcal{N}(\mathbf{r}_{T})$
is the dipole-nucleon scattering amplitude. 
The non-perturbtaive part $N(\mathbf{r}_{T},\mathbf{b}_{T})$ is chosen based on Ref.\citep{Xing:2020hwh}, which has been shown to accurately describe the data from $\rho^{0}$  production in Au-Au collisions \citep{STAR:2022wfe}.
Following Ref. \citep{Xing:2020hwh},
$B_{p}$ is chosen as $4\textrm{ GeV}{}^{-2}$, $\mathcal{N}(\mathbf{r}_{T})$
is given by the modified IP saturation model \citep{Lappi:2010dd,Lappi:2013am},
\begin{equation}
\mathcal{N}(\mathbf{r}_{T})=1-\exp\left[-r_{T}^{2}G(x_{g},\mathbf{r}_{T})\right],
\end{equation}
where $G(x_{g},\mathbf{r}_{T})$ is the gluon distribution function
with $x_{g}$ being the gluon's momentum fraction. For convenience,
one usually assumes that $G(x_{g},\mathbf{r}_{T})$ is approximately
independent of $\mathbf{r}_{T}$, so $G(x_{g},\mathbf{r}_{T})$ can
be further parameterized as $G(x_{g})=(1/4)(x_{0}/x_{g})^{\lambda_{\textrm{GBW}}}$
in the Golec-Biernat and W\"usthoff (GBW) model \citep{Golec-Biernat:1998zce,Golec-Biernat:1999qor}.
The parameters $x_{0}=3\times10^{-4}$ and $\lambda_{\textrm{GBW}}=0.29$
are determined by fitting to HERA data \citep{Kowalski:2006hc}. 
We note that an updated set of parameters for the GBW model can also be adopted as $\lambda_{\textrm{GBW}} = 0.25$ and $x_0 = 0.4 \times 10^{-4}$, as suggested by Ref.~\cite{Golec-Biernat:2017lfv}. However, we continue to use the original parameters in Refs.~\cite{Xing:2020hwh}, which have successfully explained the observations in STAR experiments.

Ref.~\cite{Kowalski:2003hm} highlights that the lumpy nucleus model is more suitable for light nuclei than our approximation in Eq.~(\ref{eq:dipole_amplitude}). However, considering our focus on heavy nuclei and the computational resource limitations, we will continue to use the approximation in Eq.~(\ref{eq:dipole_amplitude}).

The nuclear thickness function $T_{A}(\mathbf{b}_{T})$ is integrated
nuclear mass density distribution $\rho(z,\mathbf{b}_{T})$ over the
beam direction $z$,
\begin{equation}
T_{A}(\mathbf{b}_{T})=\int dz\rho(z,\mathbf{b}_{T}).\label{eq:thickness_function_01}
\end{equation}
The mass density distribution is often taken as the Woods-Saxon (WS)
distribution in spherical coordinates,
\begin{equation}
\rho(r,\theta,\phi)=\frac{\rho_{0}}{1+\exp\{[r-R_{0}(\theta,\phi)]/a\}},\label{eq:WS}
\end{equation}
where $a$ is the skin width and $R_{0}(\theta,\phi)$ is the nuclear
radius with angular dependence characterizing the nuclear deformation.
Considering that the nucleus is the axisymmetric, $R_{0}(\theta,\phi)$
should be independent of $\phi$, so it is denoted as $R_{0}(\theta)$
and can be expanded by the spherical harmonic functions $Y_{l,m}(\theta,\phi)$
with $m=0$, 
\begin{equation}
R_{0}(\theta)=R\left[1+\beta_{2}Y_{2,0}(\theta)+\beta_{3}Y_{3,0}(\theta)+...\right],
\end{equation}
where all $Y_{l0}(\theta)\equiv Y_{l0}(\theta,\phi)$ are independent
of the azimuthal angle so we suppressed $\phi$, $\beta_{i}$ are
parameters and $R$ is the averaged nuclear radius. 

In this study, we aim to learn nuclear deformation effects in isobaric
collisions. 
The WS parameters
$R$ and $a$ are determined by matching the WS distribution to the results obtained from density functional theory (DFT) calculations \citep{Xu:2021qjw,Xu:2021vpn}. 
Unfortunately,  the deformation parameters $\beta_i$ cannot be well described by calculations based
on the density functional theory (DFT), although some efforts have
been made recently \citep{Rong:2022qez}. On the other hand, the flow observables measured in relativistic isobaric
collisions imply a large quadrupole deformation for Ru ($\beta_{{\rm 2,Ru}}=0.16$)
and a large octupole deformation for Zr ($\beta_{3,Zr}=0.20$) \citep{STAR:2021mii,Zhang:2021kxj}.
Therefore,we have chosen to adopt the values for $\beta_i$  as reported in Refs. \citep{STAR:2021mii,Zhang:2021kxj}. 
The corresponding parameters in the spherical and deformed cases are
listed in Tab. \ref{tab:Parameters}. The differential cross sections
in this study are computed by the ZMCintegral package \citep{Zhang_2020cpc_zw,Wu:2019tsf}
(also see Refs. \citep{Zhang:2019uor,Zhang:2022lje} for other applications
of the package).

\begin{figure*}
\centering\includegraphics[scale=0.36]{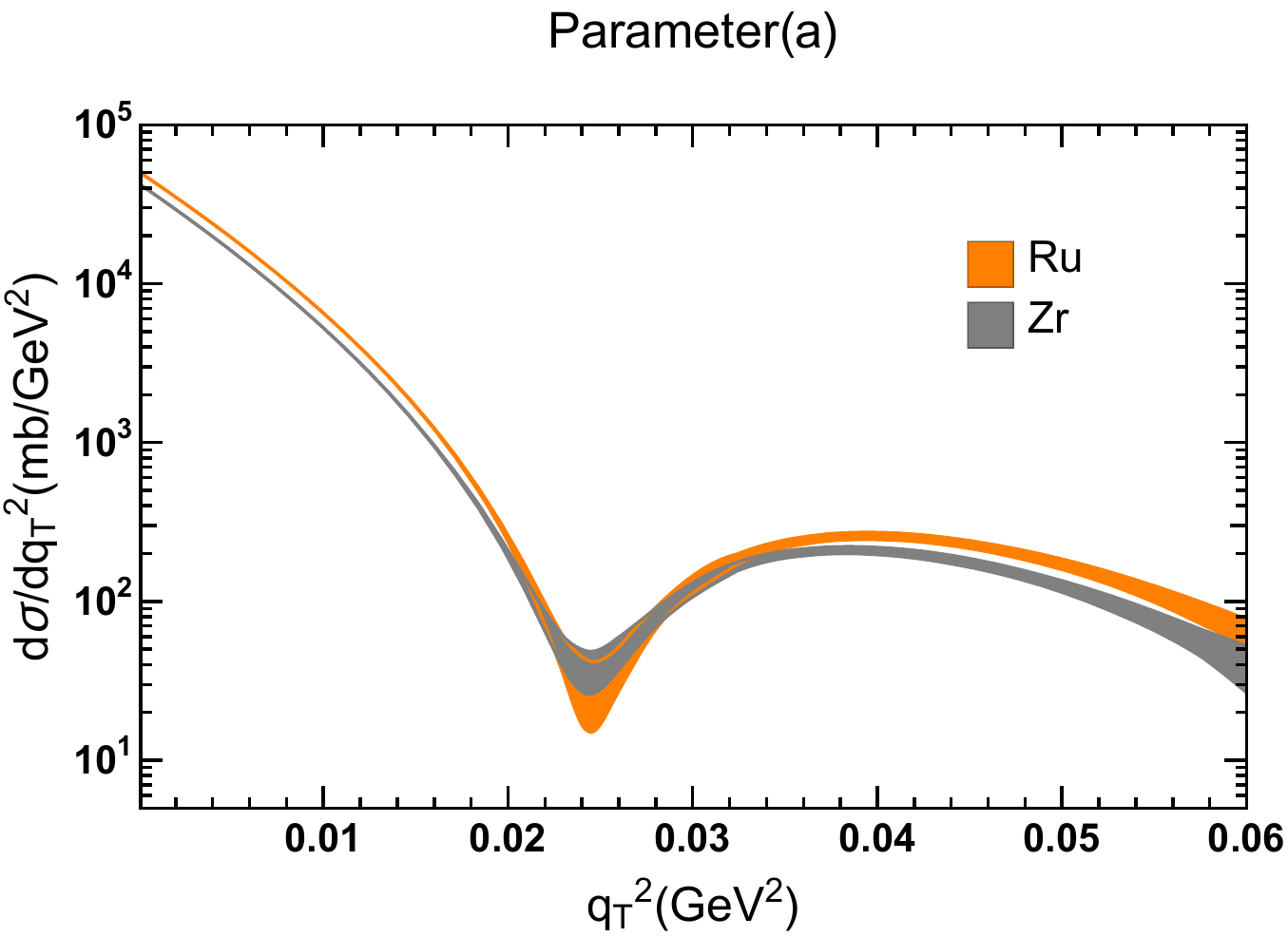}$\quad$\includegraphics[scale=0.36]{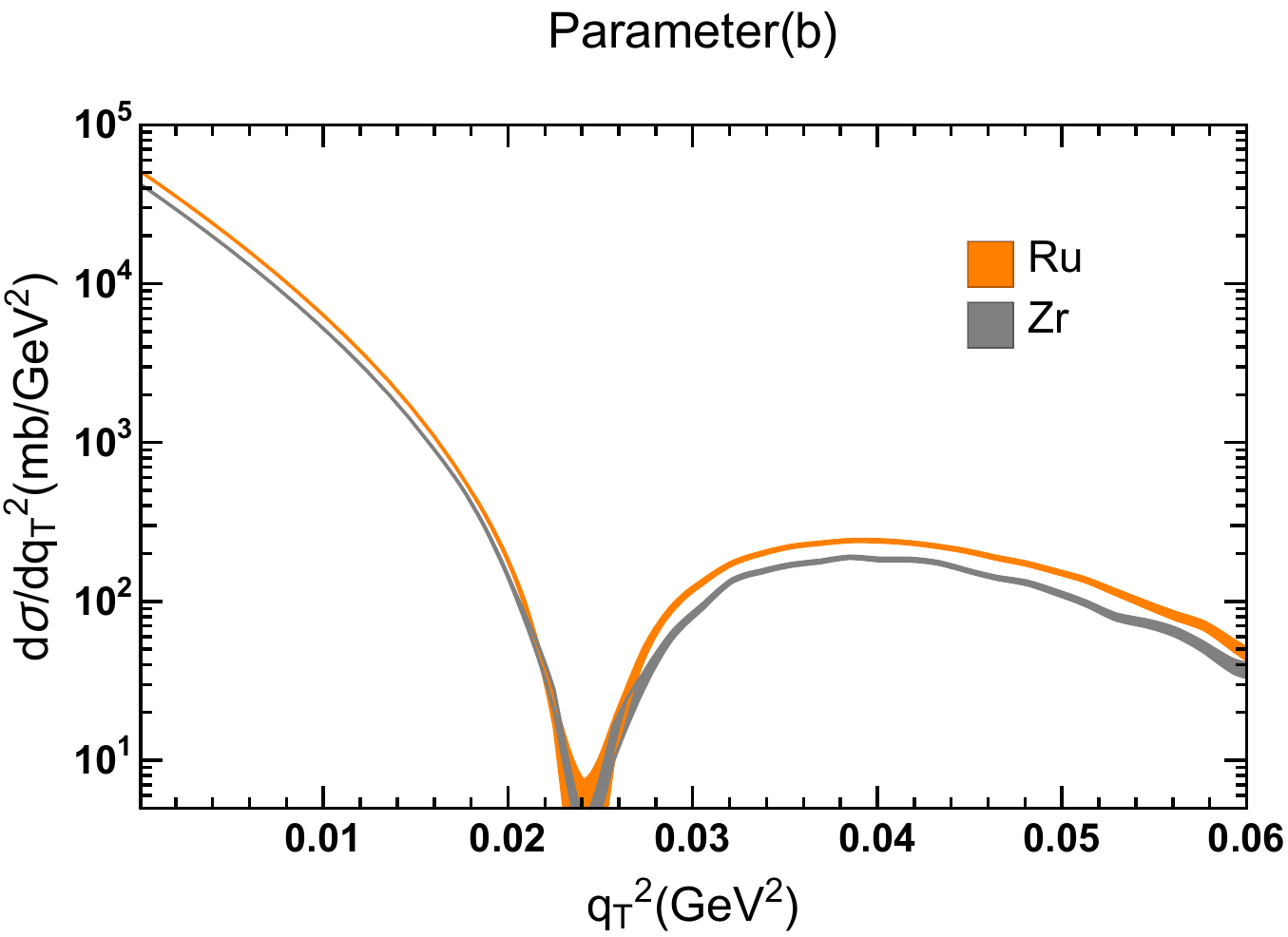}\caption{The differential cross sections as functions of transverse momentum
$q_{T}$ for $\rho^{0}$ meson (a) with and (b) without deformation
effects in Ru+Ru and Zr+Zr collisions. The parameter values are listed
in Table \ref{tab:Parameters} for deformed and spherical nuclei respectively.
The orange and gray shaded areas represent the results in Ru+Ru and
Zr+Zr collisions with estimated errors from numerical calculations,
respectively. \protect\label{fig:Pt}}
\end{figure*}

\begin{figure}
\begin{centering}
\includegraphics[scale=0.28]{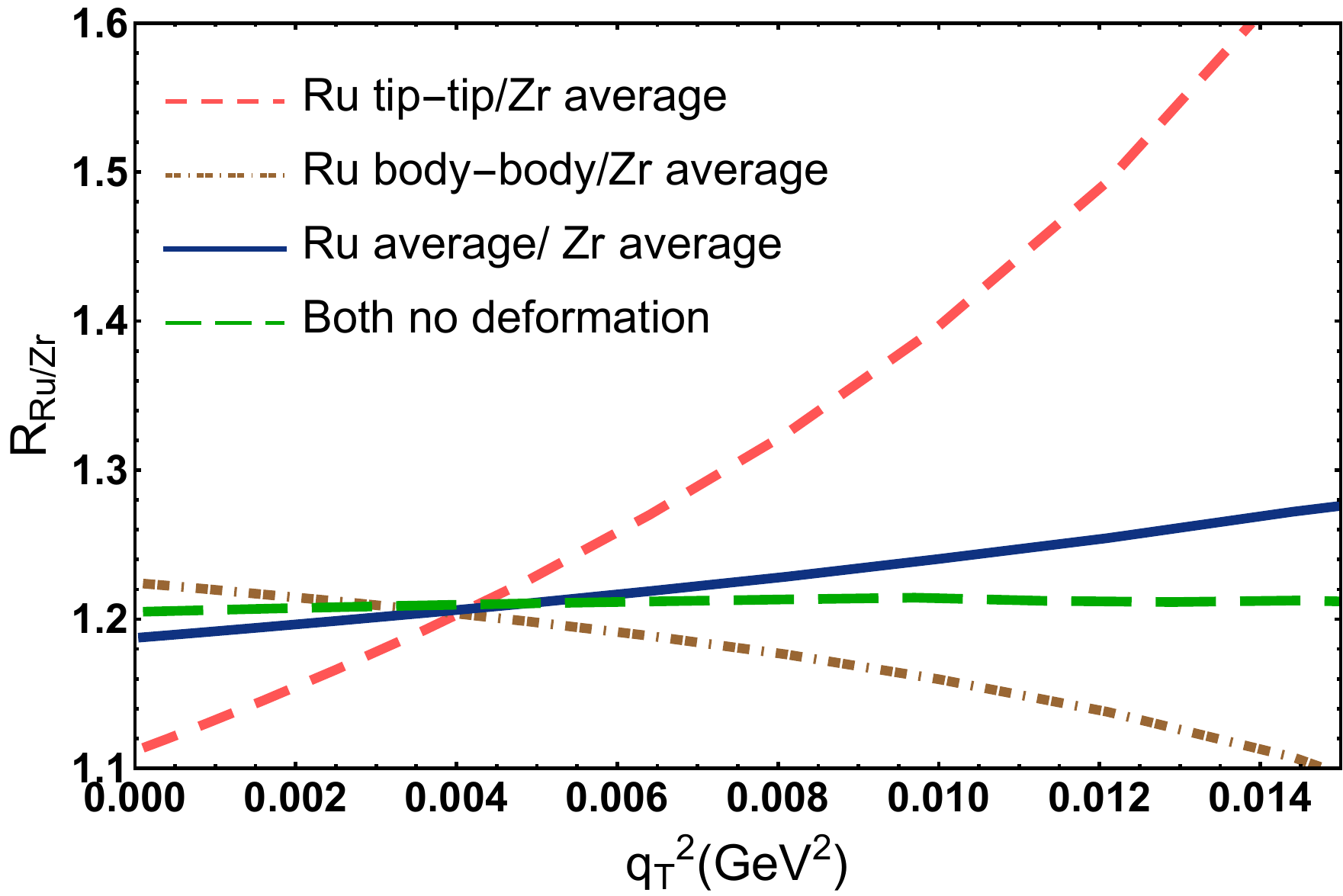}
\par\end{centering}
\centering{}\caption{The ratios of transverse momentum spectra in isobaric collisions.
The blue solid and green dashed lines represent the results with and
without nuclear deformation effects, respectively. The brown dash-dotted
and red dashed lines correspond to the results from body-body and
tip-tip configurations in Ru+Ru collisions, respectively. The parameters
for the mass density distribution are listed in Table \ref{tab:Parameters}.
\protect\label{fig:ratio}}
\end{figure}

\section{transverse momentum spectra with deformation effects }

\label{sec:distribution} We show the differential cross section $d\sigma/dq_{T}^{2}$
with and without nuclear deformation effects in Figs. \ref{fig:Pt}
(a) and (b), respectively. A notable feature is the distinct minimum
in the differential cross section at $q_{T}^{2}\simeq0.024$ $\textrm{GeV}^{2}$,
signifying diffraction characteristics of nuclear collisions. This
signature is consistent with diffraction phenomena extensively discussed
in high-energy physics \citep{Kovchegov:2012mbw,Good:1960ba,Goulianos:1982vk}.
Furthermore, the deformation effects are observed to elevate the local
minima of the cross section.

As mentioned in our theoretical framework, various models and parameters can be chosen. Indeed, we have found that the absolute values of the differential cross sections for the two nuclei are modified when different models or parameters are selected.Therefore, 

rather than analyzing the differential cross sections for Ru and Zr
independently, their ratio can yield additional insights into nuclear
structure. In the investigation of dilepton photon production in isobaric
collisions by some of us \citep{Lin:2022flv}, it is found that the
distributions of charge and mass density for Ru and Zr can lead to
the ratios of transverse momenta, invariant masses, and azimuthal
angle distributions for these nuclei that are less than the fourth
power of their charge number ratio $(Z_{\mathrm{Ru}}/Z_{\mathrm{Zr}})^{4}=(44/40)^{4}$.
However, in the present study, upon disregarding the difference in
mass distribution between Ru and Zr, the differential cross section
ratio 
\begin{equation}
R_{\mathrm{Ru/Zr}}=\frac{d\sigma_{\mathrm{Ru}}/dq_{T}^{2}}{d\sigma_{\mathrm{Zr}}/dq_{T}^{2}},
\end{equation}
is expected to scale with the square of the charge number ratio $(Z_{\mathrm{Ru}}/Z_{\mathrm{Zr}})^{2}=(44/40)^{2}$.

Surprisingly, we observe a dramatic effect from nuclear deformation
in Fig. \ref{fig:ratio}. As depicted by the green dashed line in
Fig. \ref{fig:ratio}, the ratio is $(44/40)^{2}\approx1.21$ and
is independent of $q_{T}^{2}$ in absence of nuclear deformation.
When nuclear deformation is considered, as depicted by the blue solid
line, the ratio exhibits an approximately linear increase with $q_{T}^{2}$
in the small $q_{T}^{2}$ region, $q_{T}^{2}\lesssim$ 0.015 GeV$^{2}$.
This behavior is consistent with experimental data \citep{JieZhao:2023}.
When $q_{T}^{2}\gtrsim0.015\textrm{ GeV}^{2}$, as shown in Fig. \ref{fig:Pt}(a),
we find that numerical errors increase significantly when nuclear
deformation is considered. Consequently, to maintain a sufficient
accuracy, we limit our discussion to the region $q_{T}^{2}\lesssim0.015$
GeV$^{2}$. To illustrate the impact of nuclear deformation on the
ratios of transverse momentum spectra, we analyze two extreme configurations
in Ru+Ru collisions in Fig. \ref{fig:ratio}: the tip-tip and body-body
collisions. Intriguingly, the cross section ratio of tip-tip Ru+Ru
collisions to average Zr+Zr collisions increases significantly with
$q_{T}^{2}$, whereas the ratio of body-body Ru+Ru collisions decreases
with $q_{T}^{2}$.

\begin{figure*}
\centering\includegraphics[scale=0.3]{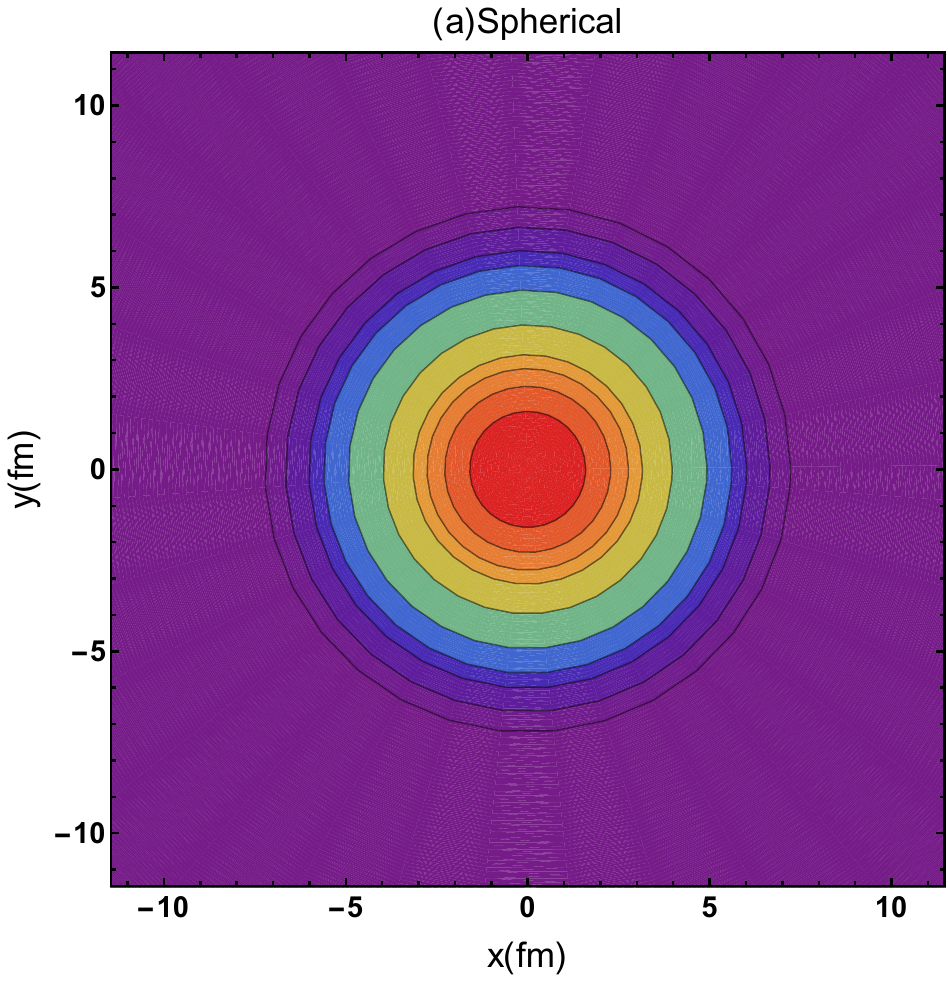}\includegraphics[scale=0.3]{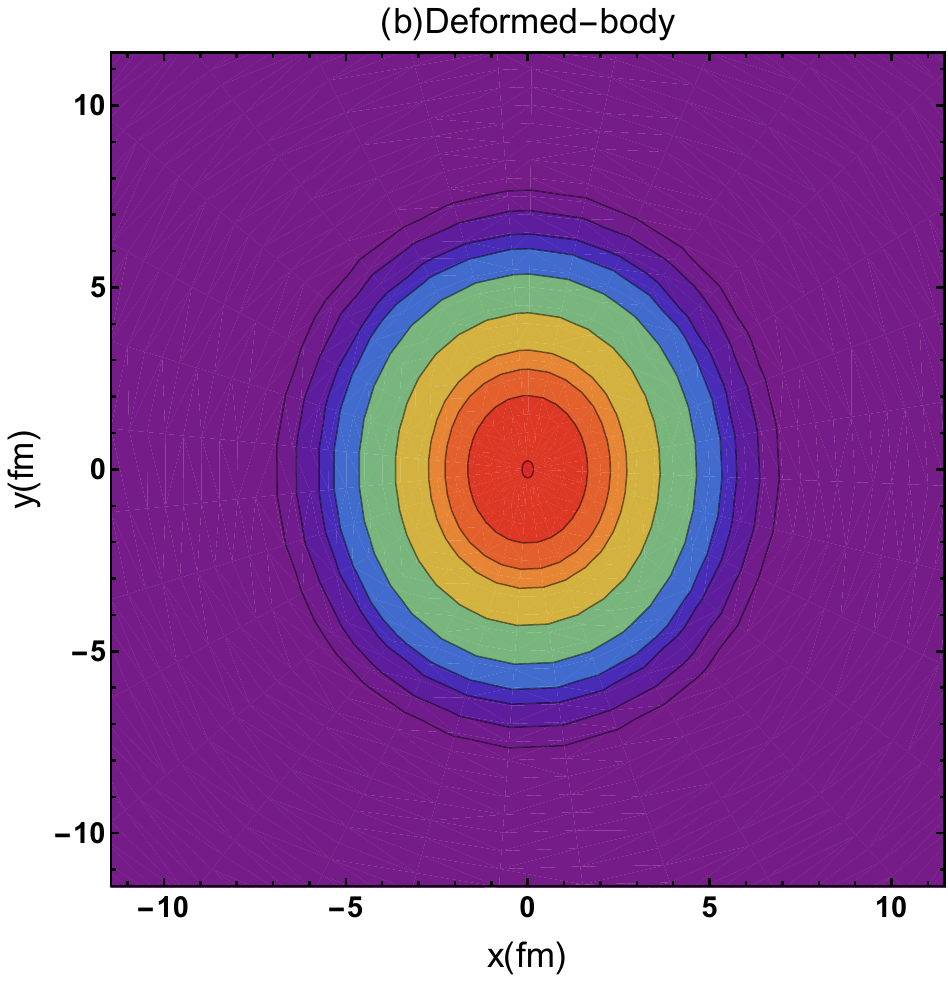}\includegraphics[scale=0.3]{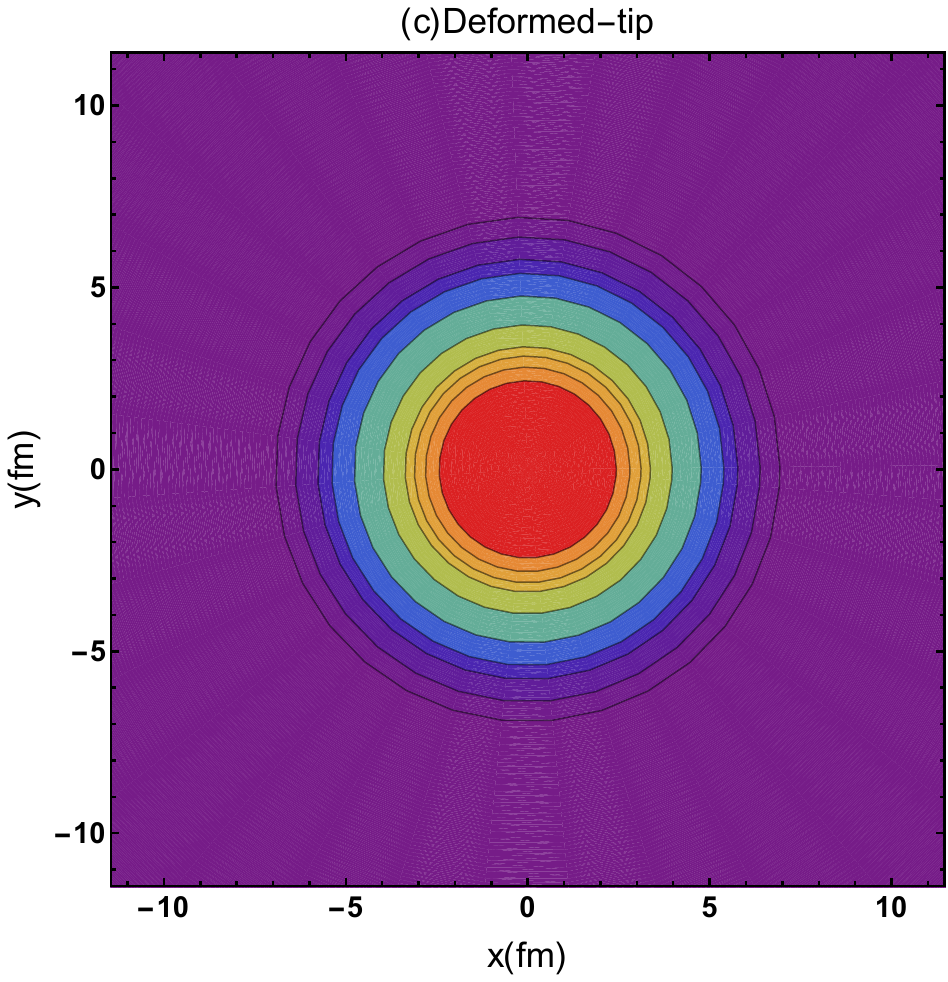}~~\includegraphics[scale=0.35]{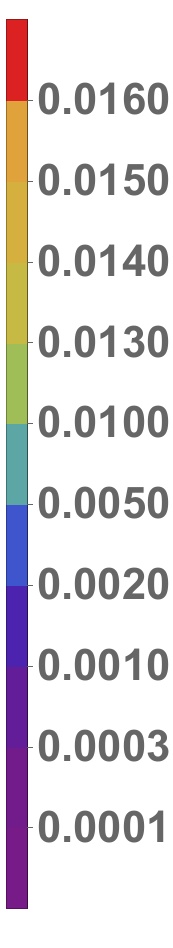}\caption{The thickness functions of the Ru nucleus in the transverse plane
for different configurations. (a) Spherrical nucleus; (b) Deformed
nucleus in body-body collisions; (c) Deformed nucleus in tip-tip collisions.
The parameters are given in Table \ref{tab:Parameters}. \protect\label{fig:Thickness}}
\end{figure*}

\begin{figure}
\centering\includegraphics[scale=0.45]{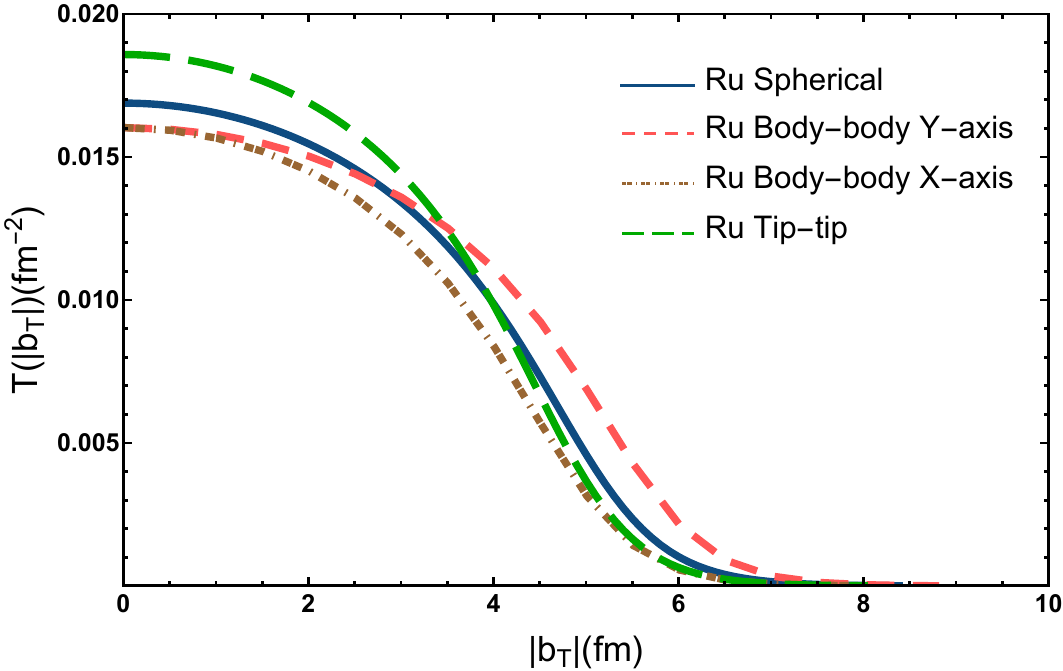}

\caption{The thickness functions of the Ru nucleus in the transverse distance
for different configurations. The blue solid line corresponds to the
spherical configuration, the red dashed line to the configuration
in body-body collisions in the y-axis, the brown dash-dotted line
to the configuration in body-body collisions in the x-axis, and the
green dashed line to the configuration in tip-tip collisions. \protect\label{fig:Thickness-1}}
\end{figure}

The nuclear deformation is manifested as variations in shape profiles
of colliding nuclei. These variations can be quantitatively analyzed
using the thickness function in Eq. (\ref{eq:thickness_function_01})
which are shown in Figs. \ref{fig:Thickness} and \ref{fig:Thickness-1}.

Now we offer a simple intuitive explanation for nuclear deformation
effects observed in the calculation. To this end, the thickness function
is parameterized by a Gaussian distribution of $\mathbf{b}_{T}$
\begin{equation}
T_{A}(\mathbf{b}_{T})\sim\exp\left(-\frac{\mathbf{b}_{T}^{2}}{w_{T}^{2}}\right),\label{eq:thickness_function_02}
\end{equation}
where $w_{T}$ is a parameter that can be interpreted as the effective
width of the nucleus. It is determined by 
\begin{equation}
w_{T}^{2}=\frac{\int d^{2}\mathbf{b}_{T}T_{A}(\mathbf{b}_{T})\mathbf{b}_{T}^{2}}{\int d^{2}\mathbf{b}_{T}T_{A}(\mathbf{b}_{T})},\label{eq:w_T}
\end{equation}
where $T_{A}(\mathbf{b}_{T})$ is given by (\ref{eq:thickness_function_01}).
The results of $w_{T}$ for various nuclear configurations are summarized
in Table \ref{tabw_T}.

\begin{table}
\caption{Estimates of the effective width $w_{T}$ for various configurations
by using Eq. (\ref{eq:w_T}). \protect\label{tabw_T}}

\centering{}%
\begin{tabular}{c|c|c|c|c}
\hline 
 & Ru (body) & Ru (tip) & Ru (spherical) & Zr (spherical)\tabularnewline
\hline 
$w_{T}$ (fm) & 3.628 & 3.372 & 3.544 & 3.571\tabularnewline
\hline 
\end{tabular}
\end{table}

We note that the mass density in the tip configuration is more concentrated
than in the spherical and body configuration, particularly in the
central region. Therefore, the effective width $w_{T,\textrm{tip}}^{\mathrm{Ru}}$
is smaller than $w_{T,\textrm{spherical}}^{\mathrm{Ru}}$. In the
body configuration, we encounter a divergence in effective parameters
along $x$ and $y$ axes, with the averaged $w_{T,\textrm{body}}^{\mathrm{Ru}}$,
closely mirroring the parameter of the spherical configuration, suggesting
$w_{T,\textrm{spherical}}^{\mathrm{Ru}}<w_{T,\textrm{body}}^{\mathrm{Ru}}$.
Overall, we observe the following relation, 
\begin{equation}
w_{T,\textrm{tip}}^{\mathrm{Ru}}<w_{T,\textrm{spherical}}^{\mathrm{Ru}}\approx w_{T,\textrm{spherical}}^{\mathrm{Zr}}<w_{T,\textrm{body}}^{\mathrm{Ru}}.\label{eq:weight_01}
\end{equation}
When two deformed nuclei in the shape of rugby balls collide, the
contact area and interaction dynamics change significantly depending
on whether the nuclei hit each other at their ends (tip-tip) or sides
(body-body). In tip-tip collisions, the nuclei interact over a smaller
area with higher mass densities. Conversely, in body-body collisions,
the larger contact area results in a lower mass density distribution. 

To be more specific, we consider the averaged dipole amplitude in
Eq. (\ref{eq:dipole_amplitude}), 
\begin{equation}
N(\mathbf{r}_{T},\mathbf{b}_{T})\simeq2\pi B_{p}AT_{A}(\mathbf{b}_{T})\mathcal{N}(\mathbf{r}_{T}).
\end{equation}
The scattering amplitude $\mathcal{A}$ can then be evaluated as,
\begin{eqnarray}
\mathcal{A} & \sim & A\int d^{2}\mathbf{r}_{T}\int_{0}^{1}dzB_{p}\mathcal{N}(\mathbf{r}_{T})\nonumber \\
 &  & \times\Psi^{\gamma\rightarrow q\bar{q}}(\mathbf{r}_{T},z)\Psi^{V\rightarrow q\bar{q}*}(\mathbf{r}_{T},z)\nonumber \\
 &  & \times\int d^{2}\mathbf{b}_{T}e^{i\mathbf{q}_{T}\cdot\mathbf{b}_{T}}\exp\left(-\frac{\mathbf{b}_{T}^{2}}{w_{T}^{2}}\right),
\end{eqnarray}
where, as mentioned in EPA, we have set $\boldsymbol{\Delta}_{T}\simeq\mathbf{q}_{T}$.
It is straightforward to find the $q_{T}^{2}$ dependence of $\mathcal{A}(q_{T}^{2})$
can be expressed as, 
\begin{equation}
\mathcal{A}(q_{T}^{2})\propto\exp\left(-\frac{1}{4}q_{T}^{2}w_{T}^{2}\right),
\end{equation}
Therefore, the ratio of differential cross sections for Ru+Ru and
Zr+Zr collisions is 
\begin{equation}
R_{\mathrm{Ru/Zr}}\propto e^{\delta w_{T}q_{T}^{2}},\label{eq:ratio_qT_01}
\end{equation}
with
\begin{equation}
\delta w_{T}\equiv-\frac{1}{2}\left[\left(w_{T}^{\mathrm{Ru}}\right)^{2}-\left(w_{T}^{\mathrm{Zr}}\right)^{2}\right].
\end{equation}
We have numerically checked the modification of $R_{\mathrm{Ru/Zr}}$ by using alternative models and parameters mentioned in Sec.~\ref{sec:Theoretical} and Sec.~\ref{sec:Phenomenology-setup}. The changes are almost negligible, which corroborates our analysis presented above.

Now, we can apply the Eq. (\ref{eq:ratio_qT_01}) to interpret Fig.
\ref{fig:ratio}. First, since $w_{T,\textrm{spherical}}^{\mathrm{Ru}}\approx w_{T,\textrm{spherical}}^{\mathrm{Zr}}$
as shown in Table (\ref{tabw_T}) and the ratio for two spherical
nuclei should be nearly constant, see the green line in Fig. \ref{fig:ratio}.
In realistic Zr+Zr collisions, we find that the octupole deformation
of Zr nucleus does not have a significant effect in the differential
cross section. Thus, $\delta w_{T}$ can be estimated using $w_{T,\textrm{spherical}}^{\mathrm{Zr}}$
in Table \ref{tabw_T} for simplicity. According to the inequality
(\ref{eq:weight_01}), it is straightforward to assert that the ratio
of the differential cross section of tip-tip Ru+Ru collisions over
the averaged Zr+Zr collisions increases with $q_{T}^{2}$. The slope
parameter $\delta w_{T}$ from Eq. (\ref{eq:ratio_qT_01}) is approximately
$17\textrm{ GeV}^{-2}$, which is roughly consistent with the slope
of the red dashed line in Fig. \ref{fig:ratio}. In contrast, body-body
Ru+Ru collisions exhibit opposite behavior for the ratio of the differential
cross sections with the slope parameter $\delta w_{T}\simeq-5\textrm{ GeV}^{-2}$.
This value is in close agreement with the slope of the brown dash-dotted
line in Fig. \ref{fig:ratio}. The results for realistic Ru+Ru collisions
with various configurations, corresponding to the blue solid line
in Fig. \ref{fig:ratio}, should fall between tip-tip and body-body
collisions. Therefore we see that the ratio of differential cross
sections for realistic Ru+Ru and Zr+Zr collisions can provide information
for nuclear structure. 

Before the end of this section, we would like to comment on Eq. (\ref{eq:ratio_qT_01}).
We emphasize that Eq. (\ref{eq:ratio_qT_01}) only provides a simple
understanding on the numerical results in Fig. \ref{fig:ratio}. For
realistic Ru+Ru collisions, it is insufficient to estimate the differential
cross section using the effective width $w_{T}$, which does not contain
sufficient information regarding diverse configurations of the quadrupole
deformed nucleus, distinct from the octupole deformation of the Zr
nucleus mentioned above. The rigorous treatment is to use Eq. (\ref{eq:cross_section_02})
which represents the average over the differential cross section for
all nuclear configurations. All numerical results in this work are
calculated using Eq. (\ref{eq:cross_section_02}).

\section{Application to photon-nuclear interaction in EIC }

\label{sec:EIC} We now extend our discussion on nuclear structure
effects in the EIC. We will use Eq. (\ref{eq:sub_cross_section})
for the photon-nuclear cross section instead of using Eq. (\ref{eq:cross_section_02}).
We consider nuclear targets Cu-63, Au-197, and U-238 as proposed for
the EIC. The nuclear mass densities are described by the standard
Woods-Saxon distribution (\ref{eq:WS}), with the parameters $R$,
$a$ and $\beta$ being taken from Ref. \citep{Loizides:2014vua}
and listed in Table \ref{tab:Parameters-1}.

\begin{table}
\caption{Parameters in the Woods-Saxon distribution (\ref{eq:WS}) for Cu-63,
Au-197, and U-238, taken from Refs. \citep{Loizides:2014vua}. \protect\label{tab:Parameters-1}}

\centering%
\begin{tabular}{c|c|c|c|c}
\hline 
 & $R$ (fm) & $a$ (fm) & \multirow{1}{*}{$\beta_{2}$} & $\beta_{3}$\tabularnewline
\hline 
$^{63}\textrm{Cu}$ & 4.200 & 0.596 & 0 & 0\tabularnewline
\hline 
$^{197}\textrm{Au}$ & 6.380 & 0.535 & 0 & 0\tabularnewline
\hline 
$^{238}\textrm{U}$ & 6.831 & 0.598 & 0.239 & 0\tabularnewline
\hline 
\end{tabular}
\end{table}

\begin{figure}
\centering\includegraphics[scale=0.35]{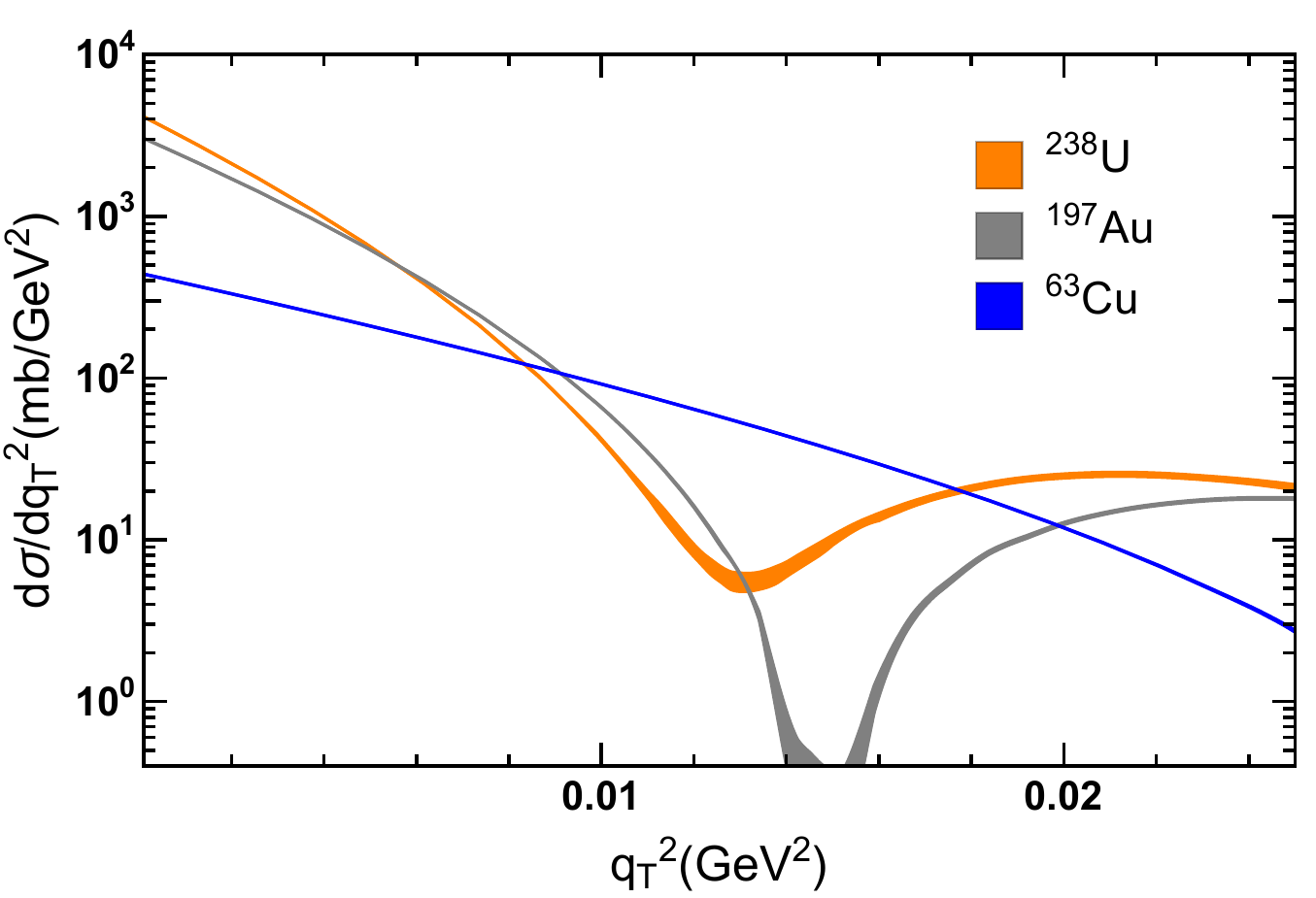}

\caption{Differential cross sections for $\gamma+A\rightarrow\rho^{0}+A$ in
the EIC as functions of $q_{T}^{2}$. The gluon momentum fraction
$x_{g}$ is set to $1.7\times10^{-3}$. The blue, gray and orange
shaded areas correspond to Cu-63, Au-197, and U-238 targets respectively,
with error estimated from numerical calculations. \protect\label{fig:EIC}}
\end{figure}

The calculated differential cross sections using Eq. (\ref{eq:sub_cross_section})
for Cu-63, Au-197, and U-238, with the typical gluon momentum fraction
$x_{g}=1.7\times10^{-3}$ \citep{Mantysaari:2023qsq}, are presented
in Fig. \ref{fig:EIC}. Substantial differences in differential cross
sections as functions of $q_{T}^{2}$ are observed for these nuclear
targets. In the region $q_{T}^{2}\lesssim0.006\textrm{ GeV}^{2}$,
we see $d\sigma_{{\rm U}}>d\sigma_{{\rm Au}}>d\sigma_{{\rm Cu}}$,
which is primarily attributed to different numbers of nucleons.

The characteristic dip behavior, associated with diffraction phenomena
in high-energy physics, is observed for all three types of nuclei,
consistent with expectation. We note that the larger the nuclear radius
$R$, the more rapidly the dip is reached as $q_{T}^{2}$ increases.
We also expect that a larger $R$ correlates with a greater $w_{T}$
in the thickness function (\ref{eq:thickness_function_02}). The decrease
behaviors of differential cross sections for three types of nuclei
before their first dips agree with our analysis of the effective width.

\section{Summary and discussion }

\label{sec:Conclusion} We have investigated the photoproduction of
$\rho^{0}$ mesons in ultraperipheral isobaric collisions of Ru+Ru
and Zr+Zr at $s_{NN}^{1/2}=$200 GeV, utilizing the dipole model with
the equivalent photon approximation. In this study, the dipole amplitude
is linked to thickness functions of nuclei, which in turn are derived
by integration of nuclear mass density distributions over the coordinate
in the beam direction. The nuclear mass density used in this work
is inspired by the DFT calculation and incorporates nuclear deformation
effects through parameters.

We calculated transverse momentum spectra in Ru+Ru and Zr+Zr collisions
and observed the characteristic dip behavior, an indicator for diffraction
processes in high-energy collisions. We further analyzed the ratio
of transverse momentum spectra for Ru+Ru and Zr+Zr collisions. In
absence of nuclear deformation, the ratio approaches $(44/40)^{2}$
approximately, whereas the inclusion of deformation effects can result
in an approximate linear increase with $q_{T}^{2}$ for $q_{T}^{2}\lesssim0.015$
$\textrm{GeV}^{2}$. This feature aligns with the trend observed in
experimental data.

We introduced a simplified Gaussian-type nuclear thickness function
in terms of the effective width $w_{T}$ and derived the dependence
of the differential cross section on $w_{T}$. By analyzing the variation
of $w_{T}$ in tip-tip and body-body collisions of Ru+Ru, we offer
an intuitive explanation for the observed behavior in ratios of transverse
momentum spectra for Ru+Ru to Zr+Zr collisions.

We also calculated the photoproduction of $\rho^{0}$ mesons in the
EIC for nuclear targets Cu-63, Au-197, and U-238. We found a significant
dependence on nuclear deformation in transverse momentum spectra.
Our results indicate that it is possible to extract nuclear deformation
parameters in ultraperipheral nuclear and electron-ion collisions.

\bigskip{}

\begin{acknowledgments}

\appendix
We would like to thank Jie Zhao, Xiao-feng Wang, Cheng Zhang, Du-xin
Zheng for helpful discussions. This work is supported in part by the
National Key Research and Development Program of China under Contract
No. 2022YFA1605500, by the Chinese Academy of Sciences (CAS) under
Grants No. YSBR-088 and by National Nature Science Foundation of China
(NSFC) under Grants No. 12075235, 12135011, 12275082, 12035006, and
12075085, 12147101.
\end{acknowledgments}

\bibliographystyle{h-physrev}
\bibliography{ref-rho-isobar}

\end{document}